\documentstyle[10pt]{article}

\def\be{\begin{equation}}
\def\ee{\end{equation}}
\def\bea{\begin{eqnarray}}
\def\eea{\end{eqnarray}}

\newcommand{\k}{\ell}
\newcommand{\dk}{d\k}
\newcommand{\kp}{{\ell}'}

\newcommand{\nlt}{n(\ell,t)}

\newcommand{\paa}{\partial}
\newcommand{\dl}{\dot{\ell}}

\newcommand{\bxi}{\bar{\xi}}

\newcommand{\dle}[1]{\label{#1}}

\newcommand{\dr}[1]{\ref{#1}}
\newcommand{\dc}[1]{~\cite{#1}}

\begin{document}

\begin{center}
{\bf EVOLUTION OF A NETWORK OF COSMIC STRING LOOPS}\footnote{Talk
given at PASCOS 98, Northeastern University, Boston, March 1998.}
\vskip 0.5cm 
{D. A. STEER}
\\
{\em{ D.A.M.T.P., Silver Street, Cambridge, CB3 9EW, U.K.\\E-mail:
D.A.Steer@damtp.cam.ac.uk} }
\end{center}

\vskip 0.7cm 

\begin{abstract}
We discuss and summarise the predictions of
 a model\dc{CKS} for the
non-equilibrium evolution of a
network of cosmic strings initially containing {\em only loops} and
{\em no infinite strings}.
The results are of interest given recent work
highlighting the problems with structure formation from the standard
cosmic string scenario\dc{Battye,Neil}.
\end{abstract}

\section{Introduction}

Cosmic strings, the analogues of vortices in superfluids,
 are one of a number of 
topological
defects which may have formed at the GUT phase
transition\dc{ViSh}.  
As they are topologically
stable, they may have
survived until today, and due to their 
strong gravitational effects they provide
an alternative theory to inflation for explaining the
formation of structure
in the universe and the
temperature fluctuations in the CMB.
In the last year, though, 
the theory seems to have been in 
some 
trouble at least in a flat universe with $\Lambda=0$, as normalisation of
the predictions of cosmic strings to COBE was shown to leave
the spectrum of density perturbations seriously lacking in 
power on large scales\dc{Battye,Neil}.

However, the cosmic string community is not despondent and
there is renewed motivation 
to understand some very
important points
which may significantly 
alter the `standard' picture of cosmic
string formation and evolution\dc{ViSh}.  In this picture, used 
to `rule out' cosmic strings\dc{Battye}, the 
infinite strings lose energy through loop production and 
reach a scaling solution in which the energy density in
strings is a fixed fraction of the energy density of the universe and
all scales grow with the horizon.  
For example, work is underway to understand the effects 
of the lattice
both in the formation of strings\dc{Julian} and to combat the
obvious limitations of
numerical simulations on cosmic string
network evolution\dc{MarkGray}.
Attention is also turning to understand whether 
strings form at all in continuous transitions\dc{Raj}, and if so
whether their initial distribution still contains both
infinite strings and loops.

Motivation for considering the evolution of a network of cosmic
string {\em loops}, a very different network to the `standard' one
containing both infinite strings and loops, comes from
different directions.  Firstly, as a result of a lattice-free dynamical
simulation of a first order phase transition, Borrill\dc{Julian}
has argued that
there is no evidence for infinite strings, but only a population of
loops.  (This claim has not been disproved even though it has been
contested\dc{AndyJames}.)  Loop networks are also produced in
other situations\dc{bub}.  Such a network is expected to
evolve very differently from the standard one, and hence 
have different cosmological consequences;  might they be
such as to `save' cosmic strings?  Secondly,
it is interesting to focus
on the loop network alone (which, we believe, has not been done).  
For example, it is
often said that the 
infinite strings are responsible for structure formation; however, could
a loop network not survive long enough to perform the same task?
Finally, given the complicated nature of phase
transitions, we do not believe
that an initial distribution containing only loops is ruled out for
certain.  

Section \dr{sec:model} outlines the main features of our model for the
evolution of a cosmic string loop network.  Results and
discussions are given in section \dr{sec:res}.

\section{The Model}\label{sec:model}

We describe the network of relativistic cosmic string loops by
the distribution function $\nlt \dk $,
which is the average number of loops per unit volume
with physical
length between $\k$ and $\k+d\k$ at time $t$.
%
To obtain an equation for ${\paa \nlt }/{\paa t}$,
length-changing interactions must be specified.  
Firstly, $\dot{\k} \neq 0$ since the
length of loops changes though gravitational
radiation, expansion and redshift.  Secondly, loops can interact with
each other.  Assuming that partners are
exchanged when loops
intersect, the two dominant
length-changing interactions between loops
are:

$\bullet$ A loop of length $\k$ can intersect another loop of length
$\k'$ to give a loop of length $\k+\k'$.

$\bullet$ A loop of length $\k$ can self-intersect to produce two daughter
loops of length $\k'$ and $\k-\k'$.
\\
In the first case, if the initial lengths of the
two loops
are in the range $\k \rightarrow \k+d\k$ and $\k'
\rightarrow \k'+d\k'$ then the number of collisions per unit time per unit
volume is given by the number of loops with each of the
initial lengths times the probability $A(\k + \k';\k,\k') $ of the
collision occurring:
\be
A(\k + \k';\k,\k')(n(\k,t)d\k) (n(\k',t) d\k').
\dle{AAdef}
\ee
Similarly, for the second process, 
the number of daughter loops of length $\k' $ produced
from the self-intersection of a loop of length $\k \rightarrow
\k+d\k$ per unit time per unit volume is given by
\be
{B}(\k-\k',\k';\k)(n(\k,t) d\k).
\dle{BBdef}
\ee
%
Given (\dr{AAdef})-(\dr{BBdef}) and letting $H$ be the expansion
rate, the equation for ${\partial n}/{\partial
t}$ is\dc{CKS}
\bea
\lefteqn{\frac{\paa n}{\paa t} =  
-\nlt  \frac{\paa {\dl}}{\paa \k} -
\frac{\paa \nlt }{\paa \k} {\dl} + \lim_{\k \rightarrow
0^+}\left[n(\k,t) \dot{\k}
\right] \delta(\k) - 3H\nlt}
&&
\dle{fulla}
 \\
 & +&
 \int_{0}^{\k/2}A(\k;\k',\k-\k')n(\k',t)n(\k-\k',t)d\kp 
\dle{full1}
 \\
& - & \nlt   \int_{0}^{\infty} A(\k+\k';\k,\k')n(\k',t)d\kp 
\dle{full2}
\\
& + &  \int_{\k}^{\infty} B(\k,\k'-\k;\k')n(\k',t)d\kp - \nlt   \int_{0}^{\k/2}  B(\k',\k-\k';\k)d\kp 
\dle{fulleqn}.
\eea
In line
(\dr{fulla}), the first two terms can be written as the single
differential $-{\paa (n \dl)}/{\paa \k}$ which is the
net positive flux of loops in $\k$ space.  The
third term in (\dr{fulla}) guarantees that $
\nlt = 0 $ for all $t$ and $  \k < 0$,
as is required for a physical distribution of loops. 
The four terms in (\dr{full1})-(\dr{fulleqn}) are 
scattering integrals.  Line (\dr{full1}) gives 
the rate of production of loops of length $\k$ from the
intersection of two smaller loops of length $\k'$ and $\k-\k'$, and
its 
converse (\dr{full2}) gives the rate at which loops of length $\k$
disappear due to intersections with other loops.  Loops of length $\k$
may be produced as a result of a larger loop self-intersecting --- this
is the first term of (\dr{fulleqn}).  The second term is 
its counterpart which gives the
rate at which loops of length $\k$ disappear due to
self-intersection.  

In a {\em non-expanding universe} $\dl = H = 0$ so that
${\paa n}/{\paa t}$ is just given by the scattering terms
(\dr{full1})-(\dr{fulleqn}), and one can verify that the total energy
density $E \propto \int_0^{\infty} \nlt \k d\k$
is preserved as necessary\dc{CKS}.  In an {\em expanding universe}
(\dr{fulleqn}) is a non-linear and non-local integro partial-differential
equation.  For that reason in existing discussions\dc{ViSh},
the interactions between loops (lines (\dr{full1})-(\dr{fulleqn})) are
ignored leaving a very simple equation to solve.  
We will not do this below.

The final step on the way to solving  (\dr{fulleqn}) given an initial
distribution $n(\k,t_0)$ is to specify $A$ and
$B$.  This was done\dc{CKS} by assuming that the loops are
Brownian random walks on scales $ \gg \bxi(t)$, the persistence
length\dc{ACK}, and that they have many small scale kinks on a
scale $\zeta(t)$. Furthermore we assumed $\bxi$, $\zeta$ are constant in
a non-expanding universe, and $\bxi$, $\zeta$$\propto t$ in an expanding
one. 


\section{Results and Discussion}\label{sec:res}
\subsection{Non-expanding Universe}
Here the network should evolve until it reaches a stable
equilibrium distribution which has been described by
string statistical mechanics (SSM)\dc{ViSh}.  Analysis of
(\dr{fulleqn}) gives results entirely consistent with those of SSM:  we
find\dc{CKS} that as long as the constant energy density $E < E_c$, a 
critical value determined by parameters of the model
(such as $\bxi$, $\zeta$), then the loops tend to an equilibrium
distribution of the form $\nlt \sim { e^{-\beta \k}}{\k^{-5/2}}$
for $\k \gg \bxi$.  For $E>E_c$ we find that $\beta \rightarrow
0$ and non-analytic behaviour
of equation (\dr{fulleqn}) together with an extension of the model to
include infinite strings leads us to predict 
that an infinite string is formed at $\beta = 0$, as in SSM.
Though it might perhaps seem surprising that a pure
loop distribution can evolve to create infinite strings, this may indeed
happen:  if the initial 
energy density of loops is sufficiently large, we must
expect that mutual collisions will lead to a percolating
system, i.e., to infinite strings. 
%
%
\subsection{Expanding Universe}
Here analysis of (\dr{fulleqn}) has led us to
conclude\dc{CKS} that the network evolves in the following ways, depending
on the initial distribution of loops: 

$\bullet$ For small $E$, the loops
disappear both in the radiation and matter eras.

$\bullet$  For larger initial energy densities, $\beta \rightarrow 0$
in the radiation era, and we expect the following possibilities:
{\bf 1)}  The point $\beta = 0$ is reached when still in the radiation era or
conceivably in the matter era, in
which case infinite strings are formed with a scale invariant
distribution of loops.  Equation (\dr{fulleqn}) then breaks down and the
effect of infinite strings need to be included\dc{ACK}.
{\bf 2)}  If $\beta$ does not reach zero, the loops disappear
in the matter era.\\
Thus the only way in which the loop network can evolve
so as to stay as a loop network (i.e.\ not containing infinite
strings) is for the loops to disappear in the matter era.  
At present this is the only scenario which might be compatible with
data in a flat universe with $\Lambda = 0$
as it increases the power on large scales.



\section*{Acknowledgments}
I would like to thank my collaborators Ed Copeland and Tom Kibble for
their part in this long project. 
This work was supported by P.P.A.R.C.



\begin{thebibliography}{99}

\bibitem{CKS} E.J.Copeland, T.W.B.Kibble \& D.A.Steer, to appear in
{\em Phys.\ Rev.\ }D.



\bibitem{Battye} A.Albrecht, R.A.Battye \& J.Robinson,
{\em Phys.\ Rev.\ Lett.\ } {\bf 79}, 4736 (1997), and R.A.Battye, these proceedings.

\bibitem{Neil} U.Pen, U.Seljak \& N.Turok, {\em Phys.\ Rev.\ Lett.\ }
{\bf 79}, 1611 (1997).

\bibitem{ViSh}  A.Vilenkin and P.Shellard,
Cosmic Strings and Other Topological Defects (Cambridge University
Press, Cambridge, 1994).




\bibitem{Julian} J.Borrill, {\em Phys.\ Rev.\ Lett.\ } {\bf 76}, 3255
(1996).


\bibitem{MarkGray} G.Vincent, N.Antunes \& M.Hindmarsh,
{\em Phys.\ Rev.\ Lett.\ }{\bf 80}, 2277
(1998).


\bibitem{Raj} A.Rajantie, these proceedings.

\bibitem{AndyJames} J.Robinson and A.Yates,
{\em Phys.\ Rev.\  }D {\bf 54}, 5211
(1996).


\bibitem{bub} S.Digal, S.Sengupta and A.Srivastava,
{\em Phys.\ Rev.\  }D {\bf 56}, 2035
(1997).






\bibitem{ACK} D.Austin, E.J.Copeland \& T.W.B.Kibble, 
{\em Phys.\ Rev.\  }D {\bf 51}, 2499
(1995).



\end{thebibliography}
\end{document}